\documentclass[%
 prl,
 jmp,%
 amsmath,amssymb,
 reprint,%
]{revtex4-1}
\usepackage{hyperref}
\usepackage{graphicx}
\usepackage{amsmath}
\usepackage[usenames, dvipsnames]{color}

\begin{document}


\title{Matrix-element induced spin polarization of photoelectrons from bulk bands } 



\author{Dmitry Vasilyev}
\affiliation{Institut f{\"{u}}r Physik, Johannes-Gutenberg-Universit{\"{a}}t, Staudingerweg 7, 55128 Mainz, Germany}

\author{Katerina Medjanik}
\affiliation{Institut f{\"{u}}r Physik, Johannes-Gutenberg-Universit{\"{a}}t, Staudingerweg 7, 55128 Mainz, Germany}

\author{Sergey Babenkov}
\affiliation{Institut f{\"{u}}r Physik, Johannes-Gutenberg-Universit{\"{a}}t, Staudingerweg 7, 55128 Mainz, Germany}

\author{Martin Ellguth}
\affiliation{Institut f{\"{u}}r Physik, Johannes-Gutenberg-Universit{\"{a}}t, Staudingerweg 7, 55128 Mainz, Germany}

\author{Gerd Sch{\"{o}}nhense}
\affiliation{Institut f{\"{u}}r Physik, Johannes-Gutenberg-Universit{\"{a}}t, Staudingerweg 7, 55128 Mainz, Germany}

\author{Hans-Joachim Elmers}
\affiliation{Institut f{\"{u}}r Physik, Johannes-Gutenberg-Universit{\"{a}}t, Staudingerweg 7, 55128 Mainz, Germany}


\date{\today}

\begin{abstract}
Angular- or $k$-resolved photoelectron spectroscopy in the soft X-ray range gives access to the bulk electronic structure of materials. Here this technique is extended to the spin degree of freedom. A non-magnetic material (tungsten) was chosen, in order to exclude any initial-state spin polarization from exchange-split bands. By measurement of two spin-polarizations for different light helicity, one can distinguish between contributions of optical spin-orientation by circularly-polarized X-rays (“Fano component”) and a second component originating from interference of final-state partial waves. Both phenomena have so far only been observed at low photon energies. Their detection in the X-ray range is a milestone on the way to the "complete experiment" in bulk photoemission. 

\end{abstract}


\maketitle 

Spin-resolved photoelectron spectroscopy has a long tradition, see e.g. textbooks for work in the gas phase \cite{Kessler} and for solids \cite{Kirschnerbook, Federbook}. After pioneering work at ETH Zurich on spin polarized electrons from ferromagnetic materials \cite{Baenninger} and optical spin orientation in GaAs \cite{GaAs}, angular-resolved photoelectron spectroscopy (ARPES) with spin-resolution developed rapidly and became a powerful tool for analyzing exchange-split bands in ferromagnets \cite{Kisker}. The method was also used for symmetry-resolved band mapping of non-magnetic metals \cite{Eyers} and adsorbates \cite{Sch1986}. In the last decade, Spin-ARPES activities were strongly intensified by the discovery of topological materials with special spin textures \cite{Hsieh,Xu,Bradlyn}.

In the X-ray range only a few spin-resolved core level measurements at ferromagnets have been performed  using lab sources \cite{Roth,Klebanoff,Johnson} and hard X-rays \cite{Stryganyuk,Ueda}. In the valence range two angle-integrating measurements have been published \cite{Gloskowskii,Kozina}; in both experiments the spin signal was close to the detection limit. The low photoemission cross sections in the X-ray range so far were prohibitive for $k$-resolved spin measurements. On the other hand, angle-resolving \cite{Gray} and $k$-resolving  \cite{Medjanik} spectroscopy have revealed the power of X-ray ARPES to study the electronic structure deep in the bulk of a material, excluding surface effects. As the bulk density of states and the Fermi surface are responsible for practically all transport and thermodynamic phenomena, detailed information on the true bulk electronic structure is mandatory for basic materials research and materials tailoring. It would be highly desirable to include the spin degree of freedom in the information content of bulk-sensitive photoemission.

Previous measurements at low energies have  demonstrated that multichannel spin detection using imaging spin filters \cite{Kolbe, Tusche} can increase the effective figure-of-merit by orders of magnitude. In the present work this technique was implemented in a momentum-resolving photoelectron spectrometer, thus overcoming the count-rate limitation of X-ray ARPES. Since the studied spin-polarization is mediated by spin-orbit interaction, we have chosen to study the bulk bands of tungsten (Z=76).

Several mechanisms may give rise to photoelectron polarization; we distinguish between initial-state, final-state and matrix-element effects. Examples for materials with initial-state spin polarization are ferromagnets with exchange-split bands or the special (ground-state) spin texture induced by the Rashba effect. Both are excluded for tungsten bulk bands; the time-reversal invariant surface state with Dirac-like spin texture  \cite{Miyamoto} was not visible at the photon energies used. 

The "final-state effect" was first described by Kirschner et al. \cite{Kirschner}. Initially unpolarized electrons excited into the upper Bloch states may acquire spin polarization when crossing the surface. The crystal surface separates the Bloch-spinor regime from the free-electron spinor regime, in which the detector is placed. Matching these relativistic wavefunctions at the boundary then may lead to a net spin polarization of the transmitted electrons, if significant spin-orbit interaction is present at the surface. In the present experiment we can rule out a significant contribution of this effect, because the kinetic energy of the photoelectrons is too high. 

The matrix-element-induced spin polarization depends on the photon polarization (since the matrix element itself depends on photon polarization). The best-studied case is the optical spin orientation by circularly polarized light (also termed Fano effect), which was initially predicted for photoemission from Cs atoms \cite{Fano}. This phenomenon is exploited in the generation of spin-polarized electrons in GaAs \cite{GaAs}. The selection rules for circularly polarized light ($\Delta m_{j}=\pm1$) lead to a population of final-state partial waves with a preferential spin orientation pointing along the photon spin $s_{\gamma}$. We term the resulting spin component along the photon spin the $Fano$ $component$. 

The matrix element gives rise to a second spin component that is oriented perpendicular to the plane spanned by photon beam and outgoing electron. Heinzmann and Dil \cite{Dil} give a detailed discussion on the nature of this component, originating from a phase-shift difference between interfering final-state partial waves. We adopt the notation $P_{\perp}$ for this perpendicular component from \cite{Dil} and earlier gas-phase work  with unpolarized \cite{Heinzmann} and linearly-polarized light \cite{SchGas}. 

In this Letter we experimentally demonstrate for the  first time a non-vanishing spin-polarization of photoelectrons excited by soft X-rays from initial bulk states of a non-magnetic centrosymmetric material by the example of the body-centered cubic metal W. Exploiting the circular polarization of the incident light beam, we distinguish two different effects being responsible for the spin polarization. Both effects are caused by the spin-orbit interaction: Optical spin orientation (Fano effect) and final state interference effect, leading to  $P_{\perp}$.

\begin{figure}[h!]
\includegraphics[width=0.5\textwidth]{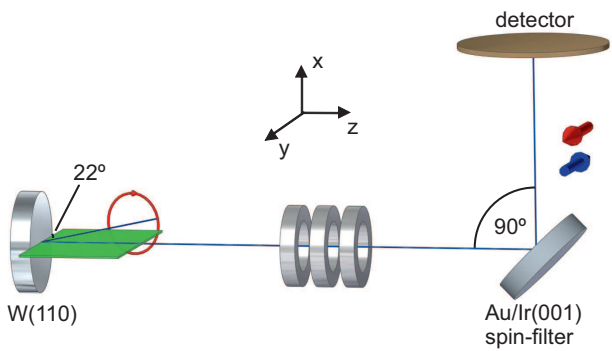}
\caption{\label{fig1}Experimental sketch of the experiment}
\end{figure}

The photoemission experiments were performed at beamline P04 of the PETRA III synchrotron center (DESY, Hamburg), providing almost completely (\textgreater 95\%) circularly polarized light in the soft X-ray regime. Here we show data taken at a photon energy of 447 eV and an incidence angle of $22^{\circ}$ with respect to the film plane. The geometry of the setup is sketched in Fig. \hyperref[fig1]{\ref{fig1}}. The plane  of photon incidence is the $yz$-plane, which coincides with the $\bar{\Gamma}$-$\mathrm{\bar{H}}$ azimuth of the crystal surface. For the detection of photoelectrons we have used time-of-flight (ToF) momentum microscopy with imaging spin filter: for details, see \cite{Schon2017,1}. The method allows detecting the photoemission intensity $I$($E_{B}$,$k_{x}$,$k_{y}$) as a function of momentum components $k_{x}$ and $k_{y}$ (parallel to the sample surface), and binding energy $E_{B}$. $k_{x}$ and $k_{y}$ are recorded by full-field $k$-imaging, ToF dispersion simultaneously resolves an energy interval of several eV width. The photon energy of 447 eV results in a $k_{z}$ value of 3.99 $G_{110}$ \cite{Medjanik}, corresponding to a cut through the $\Gamma$-$\mathrm{N}$-$\mathrm{H}$ plane in the three-dimensional Brillouin zone (BZ). 

Spin-resolution is achieved by combining full-field $k$-imaging and ToF energy recording with an imaging spin filter resulting in a very high data acquisition efficiency. It's working principle based on spin-dependent low-energy electron diffraction from a single-crystalline surface. Here, we use a pseudomorphic Au monolayer on Ir (001) as the spin-filter surface \cite{Vasilyevspinmirror}. This surface was chosen due to the high spin-sensitivity (70\%) and incredibly long lifetime (more than 12 months in ultra-high vacuum), furthermore it has a rather high effective surface Debye temperature, which provides a high efficiency at room temperature. We performed measurements at the two scattering energies of 10 eV and 11.75 eV with opposite scattering asymmetry.

We measure four photoelectron intensities,  $I_{h,l}^{+,-}$, where the upper index corresponds to the light helicity (“+” for $\sigma^+$ light and “–“ for $\sigma^-$ light) and the lower index corresponds to the spin detector scattering energy (h - high energy point 11.75 eV and l – low energy point 10 eV).

For further processing, we consider the measured reflectivity $R_{h,l}$ and the Sherman function $S_{l}=-S_{h}=S=0.7$ in order to obtain the four intensities 
\begin{equation} 
I_{up,down}^{+,-}=I_{h,l}^{+,-}/R_{h,l}S 
\end{equation}
Up (down) refers to the photocurrent for electrons with spin up (down) referring to the spin-quantization axis ($y$). One can define four independent linear combinations of these intensities. 

(i) The cumulative intensity
\begin{equation} 
I_0= I_{up}^{+} + I_{down}^{+} + I_{up}^{-} + I_{down}^{-}
\end{equation}
serves as the normalization factor in order to obtain the polarization or asymmetry values.

(ii) The spin polarization value
\begin{equation} 
P_{\perp}I_0=I_{up}^{+}-I_{down}^{+}+I_{up}^{-}-I_{down}^{-}
\end{equation}
averages out the effect of the light helicity and therefore captures the contribution arising due to final-state interference. As mentioned above, in our case of a centrosymmetric crystal and excluding interface effects, the initial state spin polarization vanishes. Since the $y-z$ plane coincides with a crystal mirror plane, the $y$-component of the spin polarization and the circular polarization simultaneously change sign upon mirroring the experiment at the $y-z$ plane. Therefore, this symmetry dictates that $P$($k_{x}$)= -$P$(-$k_{x}$). Neglecting the linear dichroism, which is suppressed to a large extent by the normalization to $I_0$, one also obtains $P$($k_{y}$)=$P$(-$k_y$) because the $x-z$ plane also coincides with a crystal mirror plane. 

(iii) The spin polarization $P_{OO}$ caused by optical orientation  is defined as
\begin{equation} 
P_{OO}I_0=
    I_{up}^{-}-I_{down}^{-}-I_{up}^{+}+I_{down}^{+}
\end{equation}
In this linear combination contribution $P_{\perp}$ (Eq.(3)) is averaged out and one detects exclusively the effect from optical spin orientation. The symmetry consideration leads to the conditions $P$($k_x$)=$P$(-$k_x$) and $P$($k_y$)=$P$(-$k_y$). The optical orientation is a matrix-element effect governed by Clebsch-Gordan coefficients. According to the relativistic selection rules, the sign of the spin polarization reflects the double-group symmetry of the initial state (Ref.\cite{Eyers} and references therein). 

(iv) The circular dichroism texture, also referred to as the circular dichroism in the angular distribution (CDAD)

\begin{equation} A_{CDAD}I_0=I_{up}^{-}+I_{down}^{-}-I_{up}^{+}-I_{down}^{+}
    \end{equation}
measures the change of the spin-integrated intensity upon reversal of the light helicity. The symmetry consideration of the present experiment results in $A$($k_x$)= -$A$(-$k_x$) and $A$($k_y$)=$A$(-$k_y$); i.e. similar to case (ii). However, its physical origin is very different and can be understood from phase-dependent overlap integrals of initial and final states as described in \cite{Cherepkov}.  

In a first step, we separately evaluate spin polarization values of photoemitted electrons for different light helicities, in order to test the existence of a finite spin polarization. The corresponding polarization values are calculated by

\begin{equation}
P^+=\dfrac{I_h^+-I_l^+}{I_h^++I_l^+}\cdot \dfrac{1}{S};\:\: P^-=\dfrac{I_h^--I_l^-}{I_h^-+I_l^-}\cdot \dfrac{1}{S}
\label{eq1}
\end{equation}

\begin{figure}[h!]
\includegraphics[width=0.5\textwidth]{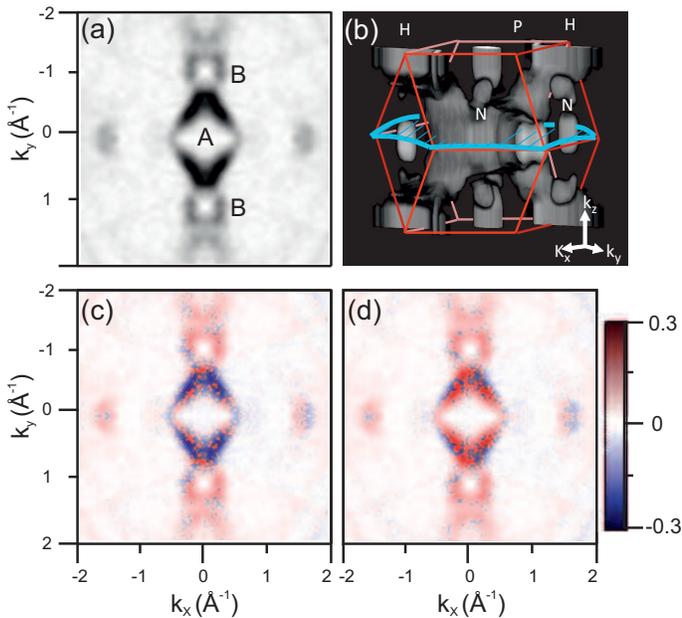}
\caption{\label{fig2}(a) experimentally measured photoemission intensity. (b) 3D Fermi surface. (c) spin-polarization for sigma minus light. (d) spin-polarization for sigma plus light. Color bar quantifies the spin-polarization.}
\end{figure}

The experimental result is shown in  Fig. \hyperref[fig2]{\ref{fig2}}, measured within 4 hours of acquisition time. One can clearly see a non-vanishing spin polarization most pronounced at the bulk states crossing the Fermi surface for both light helicities. The bulk states with sizable spin polarization correspond to the central octahedron (A) and two of the adjacent balls (B) of the Fermi surface cut close to the $\Gamma$-$N$-$H$ plane. The bulk states of the hole-like octahedron centered at the $H$-points shows almost no spin polarization. The result thus directly proves the existence of spin polarization of emitted electrons from non-polarized initial bulk states. The result for opposite helicity differ from each other and do not show any symmetries. In the following, we disentangle the spin polarization contributions from final-state interference and optical orientation.

\begin{figure}[h!]
\includegraphics[width=0.5\textwidth]{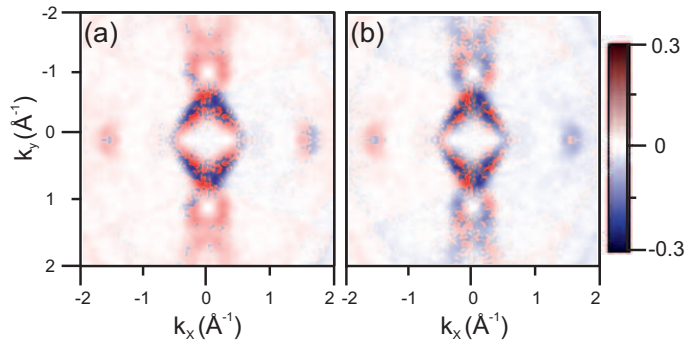}
\caption{\label{fig3}(a) Perpendicular component $P_{\perp}$. (b) $P_{\perp}$ after exploiting the symmetry conditions.}
\end{figure}

Using Eq. (3), we determine $P_{\perp}$. The result shown in Fig. 3(a) reveals a rich structure essentially reflecting the expected anti-symmetry with respect to the $y-z$ plane and symmetry with respect to the $x-z$ plane. Exploiting both conditions one obtains an artefact-free polarization map (Fig. 3(b)). The rich fine structure of $P_{\perp}$ results from the final-state wavelength (1.68\AA) being smaller than the nearest-neighbour distance (1.81\AA) and from the fact that this component results from the spin-dependent interference of outgoing partial waves.

\begin{figure}[h!]
\includegraphics[width=0.5\textwidth]{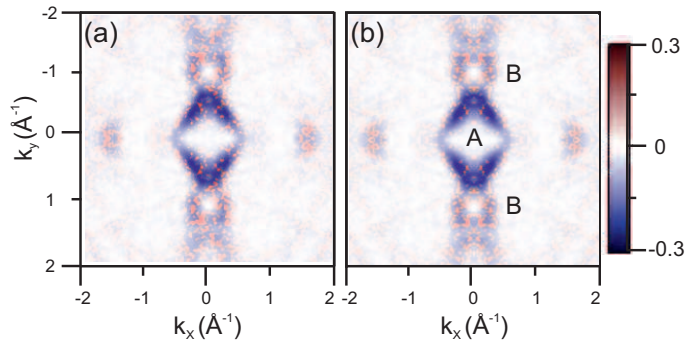}
\caption{\label{fig4}(a) Fano component $P_{OO}$ of spin-polarization.  (b)  Fano component $P_{OO}$ after exploiting the symmetry conditions.}
\end{figure}

The optical spin orientation $P_{OO}$ (Eq. (4)) depicted in Fig. 4(a) also reveals a significant spin polarization.   $P_{OO}$ shows the expected symmetry upon mirroring at the $x-z$ and $y-z$ planes. Fig. 4(b) shows the result after exploiting the symmetry conditions. $P_{OO}$ is essentially negative for the central octahedron (A) and positive for the two adjacent balls (B) reflecting the different double group symmetry of these two states. We note that this result is equivalent  to the case of the GaAs-based spin polarized electron source but using excitation at much higher photon energy.

Fig. 5(a) and the symmetrized result shown in Fig. 5(b) finally depicts the circular dichroism $A_{CDAD}$ for the high-energy working point of the spin filter. From a quasiatomic perspective, $A_{CDAD}$ is determined by the interference of different outgoing partial waves being sensitive to their phase-shift differences. As a consequence of its different origin, $A_{CDAD}$ significantly differs from $P_{\perp}$, although both have the same symmetry properties. 

\begin{figure}[h!]
\includegraphics[width=0.5\textwidth]{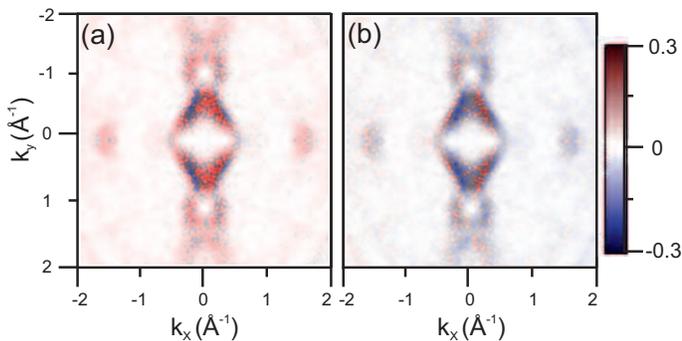}
\caption{\label{fig5}(a) CDAD asymmetry $A_{CDAD}$ measured at a scattering energy of 11.75 eV. (b) $A_{CDAD}$ after exploiting the symmetry conditions.}
\end{figure}

In conclusion, we demonstrate a finite spin polarization of photoelectrons emitted by soft X-rays from initial W(110) bulk states at the Fermi level. Spin polarization originating from optical spin orientation by circularly polarized X-rays (''Fano component'') and a contribution from interference of final-state partial waves (''component $P_{\perp}$'') are distinguished. The latter arises even for unpolarized or linearly polarized light. The spin polarization is disentangled from the circular dichroism of the photoelectron intensity (CDAD). In particular, CDAD shows a texture different from the spin polarization $P_{\perp}$, although the symmetry conditions for both quantities are identical. 

A spin polarization that does not result from optical orientation has previously been considered to involve either an interface breaking the inversion symmetry or a crystal without inversion symmetry. Therefore, the observed component $P_{\perp}$ represents a novel phenomenon. The demonstration of spin polarization of electrons excited in the soft X-ray regime, previously considered as an extremely time-consuming experiment, also paves the way to 	the analysis of the spin-polarization texture for initial bulk states in non-inversion symmetric crystals or ferromagnets.



%
%

%


\bibliography{paper}

\end{document}